# Competition between metal bonding and strain in tetragonal V$_{1-x}$M$_x$O$_2$ (M = Nb, Mo)

Jacob F. Phillips[1], Tyra C. Douglas[1], Matthew A. Davenport[1], Top B. Rawot Chhetri[1], Logan M. Whitt[1], Stephan Rosenkranz[2], Raymond Osborn[2], Matthew J. Krogstad[3], and Jared M. Allred[1]

1. Department of Chemistry and Biochemistry, University of Alabama, Tuscaloosa, AL 35487
2. Materials Science Division, Argonne National Laboratory, Lemont, IL
3. Advanced Photon Source, Argonne National Laboratory, Lemont, IL, 60439

## Abstract

Though the effects of metal dopants on the electrostructural transition of rutile VO$_2$ have been studied for many decades, there is still no consensus explanation for the observed trends. A major challenge has been to separate the impact of a dopant's size from other factors such as its electronic configuration, which stems from the difficulty in directly probing the local bonding environment around a dopant atom. This work addresses the special case of larger dopant ions by combining X-ray total scattering experiments on V$_{0.83}$Mo$_{0.17}$O$_2$ and V$_{0.89}$Nb$_{0.11}$O$_2$ single crystals with multiple Monte Carlo method models to simulate local size effects in the high-temperature tetragonal phase (R). We find that sufficiently long apical metal-oxygen bonds ($M$–O$_{ap}$) induce a strain field in the neighboring chains that locally resembles the metal-metal dimer formation present in the low-temperature distorted structure of VO$_2$ (M1). The dimer mode in the M1 structure is antisymmetric along $M$–O$_{ap}$, however, while the strain-induced pseudodimer motif is symmetric. The implied direct competition between motifs is verified experimentally. This finding provides a new mechanistic parameter toward understanding the phase transition. More generally, the work highlights how local strain fields around dopants can lead to complex distortions that are ordinarily attributed to electronic origins.

## Introduction

At 340 K, VO$_2$ undergoes a structural transition from a high temperature rutile phase to a low temperature distorted rutile monoclinic phase (M1); simultaneous to this structural transition is a transition from metallic behavior above 340 K to insulating behavior below 340 K (Goldschmidt, 1926, Morin, 1959). The relationship between these structural and electronic degrees of freedom is not well understood. The origin of the metal to insulator transition (MIT) in VO$_2$ has been the subject of much debate and various plausible mechanisms have been implicated (Budai *et al.*, 2014, Brito *et al.*, 2016, Grandi *et al.*, 2020). Both electronic and structural components of the MIT in VO$_2$ are modified by the substitution of other metals on the V site, such as Nb, Mo, Cr, Al, Ru, Fe, W, and Ti (Rawot Chhetri *et al.*, 2022, Holman *et al.*, 2009, Pouget *et al.*, 1974, Ghedira *et al.*, 1977, Gui & Cava, 2022, Kosuge & Kachi, 1976, Shibuya *et al.*, 2010, Wu *et al.*, 2015). For example, Nb, Mo, or W substitution dramatically decreases transition temperature, while Al and Cr substitution increase the transition temperature. Many times, dopants result in distorted structures as well.

The complexity of the structural phase stability in rutile phases extends beyond the electronic phase transition considerations, and it has been the subject of detailed study (Hiroi, 2015). For example, stronger metal-metal (*M-M*) bonding interactions tend to compress the chain axis (*c* lattice parameter, Fig. 1B), which in turn enlarges the $a_R$ and $b_R$ parameters to balance the electrostatics. Thus, the observed metal-oxygen octahedral shape is believed to arise from the balance between *M-M* bonding, which depends on both ion size and electron count, and coulombic interactions, which depend on ion size and charge. The resulting $c_R/a_R$ ratio has been shown to be important for predicting whether a rutile phase is susceptible to distortion or not, with VO$_2$ and isoelectronic NbO$_2$ notably lying on the verge of this line (Hiroi, 2015). MoO$_2$, by contrast, has stronger *M-M* bonding due to the extra valence electron, and so it is unambiguously in the distorted rutile regime. It has been relatedly shown that the larger ionic $M^{4+}$ radii of MoO$_2$ and NbO$_2$ should lead to greater metal octahedral distortions in the undistorted R form (Hiroi, 2015).

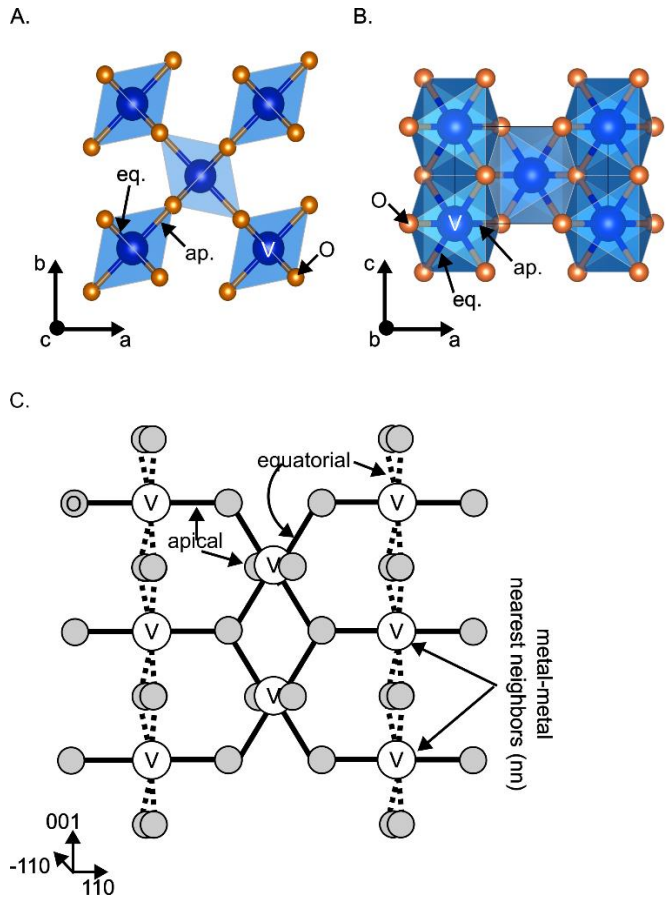

Figure 1: A. Rutile crystal structure viewed down the **c** axis.  B. Rutile crystal structure viewed down the **b** axis.  C. [-110] plane of the rutile crystal structure, viewed with a slight tilt to show out of plane oxygens.

Solid-state solutions of the three elements, V, Nb, and Mo, would thus be expected to have a complex interplay of competing structural considerations. This complexity is demonstrated in the vastly different doping responses of the *c* lattice parameter to Mo and Nb-doping, while the *a* lattice parameter response to doping is identical for the two systems (see Figure 2).

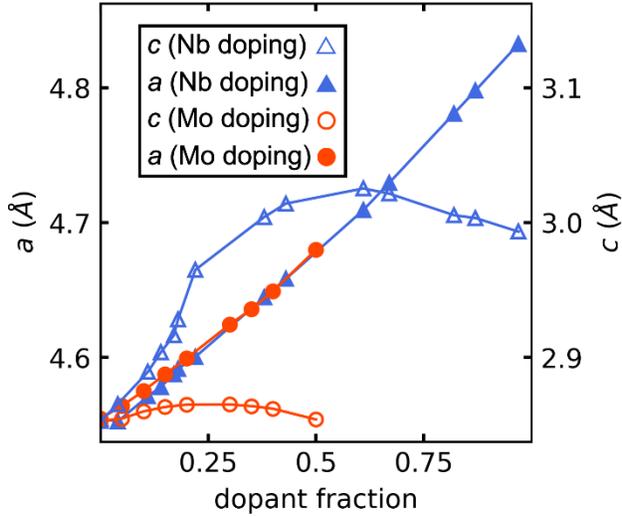

Figure 2: Lattice parameters from powder X-ray diffraction vs. dopant fraction for Mo-doped and Nb-doped VO$_2$.(Holman *et al.*, 2009, Rawot Chhetri *et al.*, 2022).

Indications of competing structural interactions are also evident in the so-called 2DM2 phases observed with Nb and Mo doping, where doping weakens structural ordering along certain crystallographic directions (Davenport *et al.*, 2021, Rawot Chhetri *et al.*, 2022). Doping destabilizes the distorted rutile phase, and at sufficient dopant percentage, only the rutile phase forms. For both Mo doping and Nb doping, this semiconducting 2D monoclinic phase is produced at low temperature instead of the insulating phase at a critical dopant fraction, then upon further doping, a metallic solid solution is produced and no transition is observed (Holman *et al.*, 2009, Davenport *et al.*, 2021, Rawot Chhetri *et al.*, 2022). The 2DM2 phase is characterized by planes of strongly correlated atomic displacements identical to the M2 phase, a distorted rutile structure observed in CrWO$_4$ and strained VO$_2$ (Vlasse *et al.*, 1976, Tselev *et al.*, 2010). Atomic displacements are also strongly correlated perpendicular to these planes, but they are prevented from ordering fully due to geometric frustration (Rawot Chhetri *et al.*, 2022). These planes of correlated displacements produce intense nets of diffuse scattering in half integer *L* reciprocal space planes. This scattering from correlated displacements is known as atomic displacement parameter (ADP) correlation scattering.

X-ray diffraction data for Mo and Nb-doped VO$_2$ often shows diffuse scattering from other types of short-range order such as size effects. Size effect scattering originates from the displacements produced when a crystallographic site is occupied by a dopant atom that is larger or smaller than the typical occupant. Size effect scattering is characterized by an odd dependence of scattering intensity with Q vector, and it occurs in the vicinity of Bragg peaks. 'Size-effect-like' scattering was noted in the high temperature phases of Nb-doped $VO_2$ at low dopant fractions (Rawot Chhetri *et al.*, 2022).

Many of the currently established findings depend on the average crystallographic structure, which typically changes continuously with metal substitution. What is not understood, however, is how the local environment of the dopant changes. That is, does the metal-oxygen geometry more closely match the local environment of the host lattice, or does it exhibit a geometry more closely related to the end member? This question is further complicated by the presence of local structural ordering phases, such as the so-called 2DM2 phase, wherein local bond formation also likely modifies the metal-oxygen geometry in some collective fashion. This is relevant to questions about the local vs. itinerant nature of valence electrons in a specific compound. Any findings in this regard comment on the electronic properties of the materials, making the questions highly relevant to the electronic structure discussion as well. For example, some interpretations rely on the assumption that Mo and Nb do not engaging in *M-M* bonding when they are introduced as dopants into $VO_2$. Such questions are usually probed indirectly, and so there is very little direct information about the local bonding environment of such dopants. This study uses simulated size effects in doped $VO_2$ single crystals to glean local structure information by comparison of simulated size effect scattering to experimental X-ray scattering data.

## Methods

Size effects were simulated for 11% Nb-doped $VO_2$ crystals using a multiple Monte Carlo (MC) procedure implemented in the computer program DISCUS (Proffen & Neder, 1997). Simulations were performed on a $50 \times 50 \times 50$ unit cell simulated crystal. Random metal site substitutions were performed with a frequency consistent with the desired dopant concentration. Atoms were displaced via an MC process, with sufficient cycles to minimize the energy of the system.. The multiple MC process was constrained with 12−6 Lennard-Jones (LJ) potentials between selected atom pairs in the crystal. The simulations reported here consider only nearest neighbor (nn) *M*-O and *M-M* pairs, where *M* = V or Nb. Each 12−6 LJ interaction potential contains 2 parameters, $\sigma$ and $\varepsilon$, which determine the distance of minimum potential and depth of the potential energy well, respectively. The minimum of the LJ potential is equal to $\sqrt[6]{2}\sigma$ and is given as $\sigma'(i-j)$; this serves as the target bond length for atoms *i* and *j*. Details on the specific MC parameters used are included in the SI.

A size effects model requires separate LJ potentials for the V and Nb sites to account for the local bonding environments of each atom type. Anisotropy in the size effects can be introduced by using separate LJ potentials for crystallographically distinct bonds. The rutile crystal structure has two distinct metal-oxygen (*M*–O) bond types: apical (ap), which are oriented toward shared corners of the octahedra between chains, and equatorial (eq), which form the shared octahedral edges along the chains (see Fig. 1). Only one nn metal-metal (*M-M*) interaction is considered, and it has three permutations: V–V, V–Nb, and Nb–Nb. This yields a maximum of seven possible LJ potentials to consider: four *M*–O type, and three *M*–*M* type. Since each LJ potential has two parameters, $\sigma(i–j)$ and $\varepsilon(i–j)$, there are a maximum of 14 definable parameters. The absolute scale of the $\varepsilon(i–j)$ parameters is set by the simulation temperature, which is held constant for all models shown. During the course of this study, several hundred models were tested using a variety of constraints in $\sigma(i–j)$ and $\varepsilon(i–j)$ values. Generally, the simulations showed very little sensitivity to relative changes in the potential depths, $\varepsilon(i–j)_i$, to the extent that all included potentials are set to be equal in the models shown here. This leaves the target bond lengths, $\sigma'(i–j)$, as the primary mode of testing bond anisotropies, resulting in 7 input parameters for the models.

To determine reasonable constraints for these models, known atomic radii for the constituent atoms in the modeled structure were referenced (Shannon, 1976). The average target Nb–O bond length, $\sigma'$(Nb–O), was derived from Shannon ionic radii of Nb in the 4+ oxidation state, and O in the 2- oxidation state, giving a value of 2.04 Å. Note that the dilute aspect of the Nb dopant means that even large changes in its effective size result in quite minor changes in the constrained $\sigma'$(V–O) target. This difference in target metal oxygen bond length between the dopant and vanadium is parameterized as the ratio between $\sigma'$(Nb–O) and $\sigma'$(V–O), $R_{MO}$. $R_{MO}$ controls the size of the dopant octahedron relative to the host vanadium octahedron.

The rutile lattice usually exhibits different *M*–O bond lengths for apical and equatorial *M*–O bond types. This is accomplished here by introducing an octahedral anisotropy term $\alpha_M = \frac{M-O\ apical\ bond\ length}{M-O\ equatorial\ bond\ length}$ for each octahedron type, with $A_M = \frac{\alpha_{Nb}}{\alpha_V}$ being a convenient value for comparison of dopant and V octahedron shape. Thus, the four *M*–O components of the octahedral anisotropy term are reduced to a single parameter, which is represented here using $A_M$. $A_M$ governs the shape of the dopant octahedron, with a value of 1 giving the same octahedron shape as in the average structure. If $A_M > 1$, the dopant octahedron is elongated along the apical bonds and compressed along the equatorial bonds relative to the average structure. If $A_M < 1$, the opposite modification is imposed.

Direct metal-metal size effects were included in some models by setting target distances between nearest neighbor metal atoms along rutile **c**. $\sigma'$(Nb–V) was constrained to have a value halfway between $\sigma'$(Nb–Nb) and $\sigma'$(V–V). These 3 target distances were parameterized as the ratio between $\sigma'$(Nb–Nb) and $\sigma'$(V–V), $R_{MM}$.

The weighted average of target apical metal-oxygen, equatorial metal-oxygen, and nearest neighbor metal-metal distances were held equal to their respective counterparts in the average structure determined previously through conventional crystallographic means (Davenport et al., 2021, Rawot Chhetri et al., 2022).

Thus, all models in this study can be described by the reduced parameters, $R_{MO}$, $A_M$, and $R_{MM}$, and their mathematical relation to the base terms is given in the SI in Table S1.

The same considerations apply to simulations of 17% Mo-doped $VO_2$ crystals. These models only have a few minor differences to distinguish them: the substitution fraction, the **c** lattice parameter, the X-ray scattering factor of the dopant site (Mo vs. Nb).

X-ray scattering from the model crystals was simulated using DISCUS. Details of the parameters used in the X-ray scattering simulation are provided in the SI.

Single crystal X-ray diffraction experiments on $Nb_{0.11}V_{0.89}O_2$ and $Mo_{0.17}V_{0.83}O_2$ were performed at the Advanced Photon Source at Argonne National Laboratory on sector 6-ID-D. Single crystals were mounted on Kapton capillaries using Duco cement for the diffuse scattering experiment. Scattering data was collected at $T = 293$ K for $Nb_{0.11}V_{0.89}O_2$ and $Mo_{0.17}V_{0.83}O_2$. The raw detector data was processed using NeXpy32 and transformed to a reciprocal space coordinate system using the Crystal Coordinate Transformation Workflow (Jennings *et al.*, Davenport *et al.*, 2021). Results from these experiments were reported in part in previous publications, and the analysis is continued in this study (Rawot Chhetri *et al.*, 2022, Davenport *et al.*, 2021).

## Results

Analysis of X-ray scattering from the high temperature phase (HT) of $Nb_{0.11}V_{0.89}O_2$ reveals unique size-effect like scattering near [0.0 2.9 3.0]. A comparison of this scattering to that of low temperature (LT) $Nb_{0.11}V_{0.89}O_2$, LT $Mo_{0.17}V_{0.83}O_2$, and HT $Mo_{0.17}V_{0.83}O_2$ is given in figure 3. In HT Nb-doped $VO_2$, the "size-effect-like" scattering appears above and below [0.0 2.9 3.0]. There is a relatively stronger peak on the -$L$ side, and a weaker peak on the +$L$ side (see Figure 3). The peak on the +$L$ side vanishes at low temperature. Mo-doped $VO_2$ has a single peak centered on [0.0 2.9 3.0], and this peak changes very little with temperature.

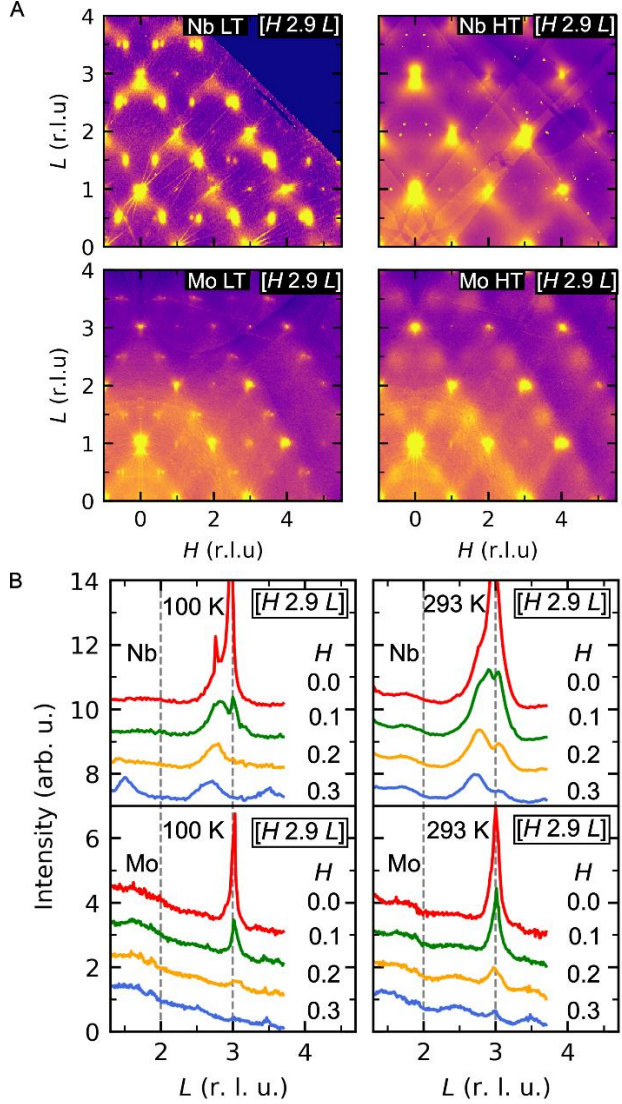

Figure 3: A. Experimental X-ray scattering from $V_{0.89}Nb_{0.11}O_2$ and $V_{0.83}Mo_{0.17}O_2$ at $T = 293$ K (HT) and $T = 100$ K (LT) in the $K = 2.9$ reciprocal space plane. B. Cuts in the vicinity of [0.0 2.9 L] are shown in panel A.

In the $L$ planes, both Nb and Mo-doped $VO_2$ have significant size effect scattering, as evidenced by asymmetric behavior of scattering intensity with the scattering vector, $Q$, along the $H(H+0.3)0$ diagonals (see Figure 4). This size effect scattering varies little with temperature and composition.

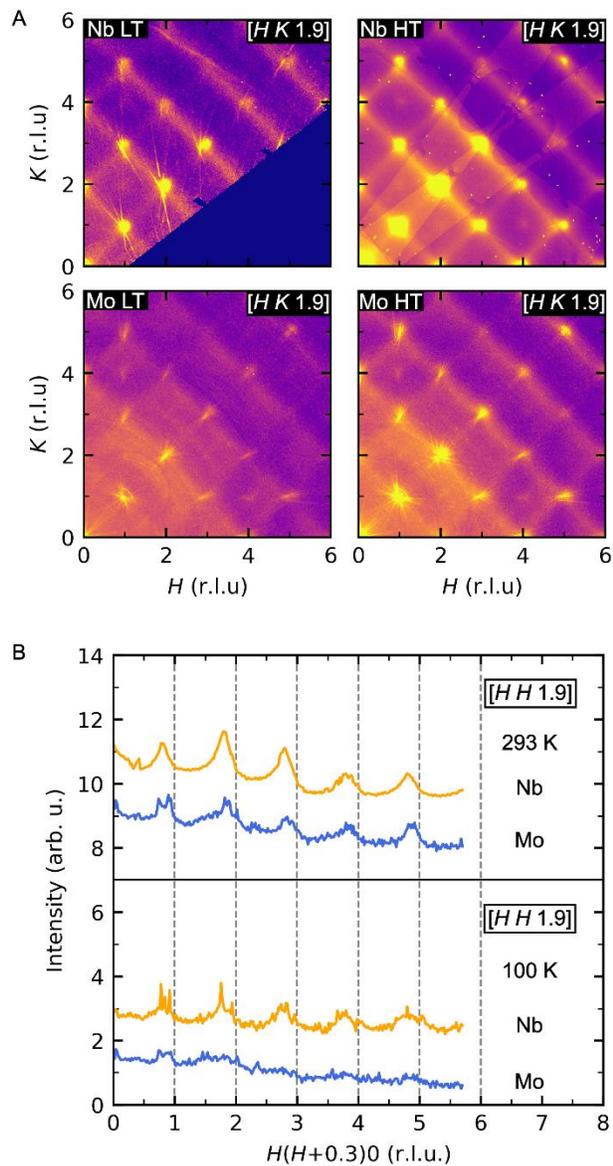

Figure 4: A. Experimental X-ray scattering from $V_{0.89}Nb_{0.11}O_2$ and $V_{0.83}Mo_{0.17}O_2$ at $T = 293$ K (HT) and $T = 100$ K (LT) in the $L = 1.9$ reciprocal space plane. B. Cuts along the $H(H + 0.3)0$ diagonal of the $L = 1.9$ reciprocal space planes shown in panel A.

Size effect models were constructed to investigate which features are needed to yield a particular chemical environment in a simulation and then show what their impacts are on the respective simulated X-ray diffuse scattering patterns. To this end, each model type is referenced by the constraints used in the LJ parameter space, rather than the specific numerical values used in a particular simulation. Since the terms usually can be applied to Nb and Mo with only trivial changes, only the Nb models are referenced unless specificity is required.

A total of 9 models are considered in this paper. The resulting simulated structure was not especially sensitive to the value of the $R_{M-O}$ parameter, so $R_{M-O}$ was fixed at a value of 1.06 for all models shown here. This sets the size of the Nb octahedron to correspond the Nb(IV) oxidation state, as discussed previously. Model crystals were simulated with metal-oxygen size effects such that $A_M$ = 0.955, 1.000, and 1.045, respectively; these models did not incorporate any direct metal-metal size effects. Another 6 model crystals were also simulated that used each of the three $A_M$ values along with adding metal-metal size effects between nearest neighbor (nn) metal atoms with $R_{MM}$ = 1.045 and 0.955, respectively. The target bond lengths for all models are given in the SI.

First, the resulting local structure of the simulated crystals will be discussed for the various models. Then, a qualitative comparison of the simulated scattering for each model to the experimental scattering for the HT phases of 11% Nb-doped $VO_2$ and 17% Mo-doped $VO_2$ will be given. In most cases, the results of the simulations using 17% Mo are impossible to distinguish from those using 11% Nb. Unless stated otherwise, it is the Nb-based simulations that are being shown.

In discussing the displacements produced in the model rutile structures, it is helpful to define a basis to describe the displacements present. The basis used here allows description of metal atom displacement in terms of 3 orthogonal displacement modes, $B_{1u}$, displacement along the rutile **c** axis, $B_{3u}$, displacement along the apical metal-oxygen bonds, and $B_{2u}$, displacement perpendicular to both $B_{1u}$ and $B_{3u}$. A depiction of this basis in the rutile structure is given in Figure 5.

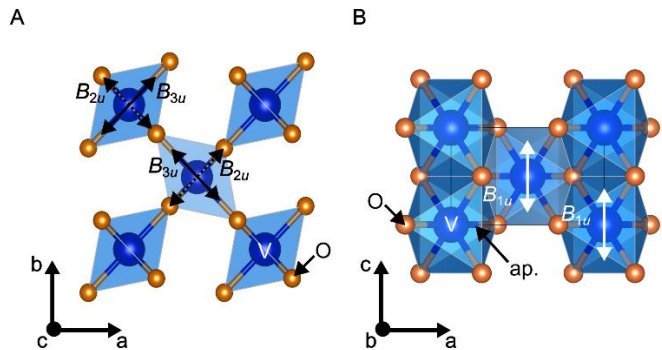

Figure 5: Metal atom displacement modes in the rutile. Dimerizing and buckling displacements are assigned to the $B_{1u}$ and $B_{3u}$ modes, respectively.

Displacement vectors for nearest neighbor (nn) metal atoms were obtained within DISCUS, and the distribution of displacements present for each atom pair type was analyzed. The distribution of $B_{1u}$ displacements for V–V nn pairs is shown in Figure 6A. It was observed that the $A_M = 1$ and $A_M > 1$ models produced a bimodal distribution of $B_{1u}$ metal displacements for all metal atom pairs, and the V–V $B_{1u}$ displacement distribution is shown as a representative example. This bimodal distribution is indicative of induced local dimer-like displacements, and it is strongly correlated with the elongation of the apical Nb–O (Mo–O) bonds relative to the average structure; a schematic depiction of the induced interchain distortion is given in Figure 6B. Note that the resulting $B_{1u}$ displacement need not be so signficant as to qualify as as a chemical bond and, and so the term 'pseudodimer' will be used when referencing the motif. This has the benefit of keeping the pseudodimer motifs distinct from the metal-metal dimer motifs present in the LT M1 phase and related structures. The discussion section provides additional context motivating this distinction. Relatedly, the pseudodimer motif is not observed in models that include metal-metal size effects of either $R_{MM} > 1$ and $R_{MM} < 1$. In such cases the extra interaction constrains the $B_{1u}$ displacement distributions away from the bimodal distribution of the pseudodimers.

The $B_{3u}$ displacements for nn metal atoms were also analyzed. Models with $R_{MM} < 1$ size effects combined with either $A_M = 1$ or $A_M < 1$ had distinct shoulders in the $B_{3u}$ displacement distribution (see Figure 6C). These shoulders are indicative of buckling displacements that alternate direction along rutile **c**, shown schematically in Figure 6D. Both the $A_M < 1$ model and the model with a combination of $A_M < 1$ and $R_{MM} > 1$ showed broadening of the $B_{3u}$ displacement distribution, indicating weak buckling displacements. For the rest of the models, the $B_{3u}$ displacements were Gaussian distributed centered on zero magnitude with no broadening or shoulders in the V–V $B_{3u}$ displacement distribution, indicating very weak or absent buckling displacements. Buckling was not evident in the $B_{3u}$ displacement distributions for V–Nb or Nb–Nb pairs, with the exception of the model with both $R_{MM} < 1$ and $A_M < 1$, which had evident buckling for all nn metal atom pairs. In summary, both increased equatorial Nb–O bond length and $R_{MM} < 1$ metal-metal interactions are required to produce significant induced buckling displacements in the simulations.

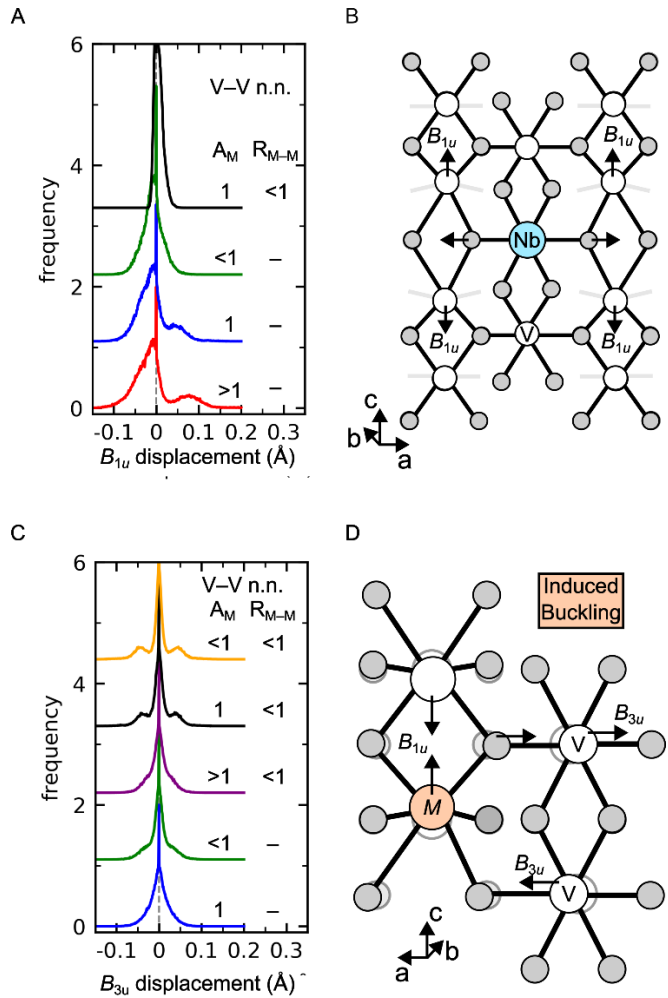

Figure 6: A. Histograms of $B_{1u}$ displacement for V–V nn in selected model structures. B. Depiction of induced dimer formation in the rutile structure. C. Histograms of $B_{3u}$ displacement for V–V nn in selected model structures. D. Depiction of induced buckling in the rutile structure.

X-ray diffuse scattering was simulated for each model in both the $K = 2.9$ plane and the $L = 1.9$ plane (see Figures 7A and 8A). [0.1 2.9 $L$] cuts were taken to probe the $L$ dependence of the simulated scattering from size effects (see Figure 7B). The $A_M > 1$ and $A_M = 1$ models produce a tailed peak both above and below [0.1 2.9 3.0]. The presence of the +$L$ component of this feature corresponds to induced pseudodimer formation in these model structures. The -$L$ portion of this feature corresponds to increased Nb–V and Nb–Nb nn distances, and all of the models, with the exception of models with $R_{MM} < 1$, have a peak at this location. The scattering in the $K = 2.9$ plane becomes more dispersed along $H$ both as $A_M$ is decreased and with the inclusion of the $R_{MM}$ parameter; this results in peaks at both even and odd $L$ in the [0.1 2.9 $L$] cut.

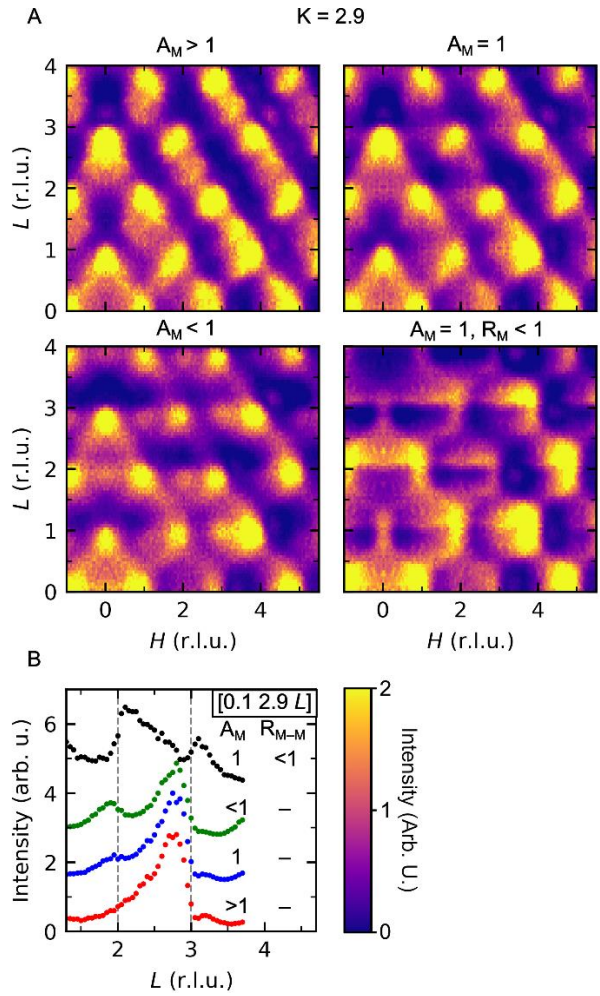

Figure 7: A. Simulated x-ray diffuse scattering in the $K = 2.9$ reciprocal space plane for selected models. B. [0.1 2.9 $L$] cuts for all reciprocal space planes shown in panel A.

Figure 8 shows X-ray diffuse scattering in the $L = 1.9$ plane simulated for selected models (panels A-D), and the respective diagonal [$H$ $H$+0.3 1.9] cuts (panel E). Many of the features in this cut are double peaks for models with $R_{MM} < 1$; this corresponds to the presence of induced buckling displacements in the model structures. Models without strong buckling displacements produce predominantly single, tailed peaks in this cut, showing significant similarity to the size effects scattering observed for the Mo and Nb-doped $VO_2$ phases (see figures 4 and 8). So, the experimental data shows no evidence of the kind of induced buckling observed in the $R_{MM} < 1$ model structures.

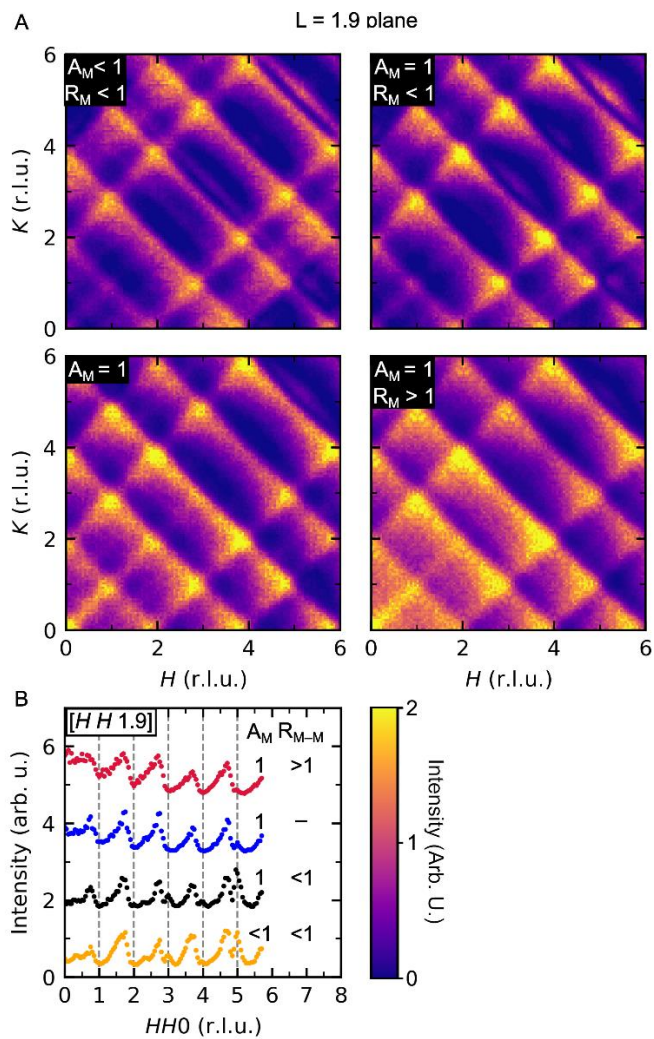

Figure 8: A. Simulated X-ray diffuse scattering in the $L = 1.9$ reciprocal space plane for selected models. B. Cuts along the $[H\ (H + 0.3)\ 0]$ diagonal of the $L = 1.9$ reciprocal space planes shown in panel A.

## Discussion

The simulated scattering from the size effect models for $Nb_{0.11}V_{0.89}O_2$ is consistent with the presence of size effects in both $Nb_{0.11}V_{0.89}O_2$ and $Mo_{0.17}V_{0.83}O_2$. The double peak feature around $[0.1\ 2.9\ 3.0]$ produced by the $A_M > 1$ and $A_M = 1$ models bears a strong resemblance to the corresponding scattering for high temperature $V_{0.89}Nb_{0.11}O_2$ (see Figures 3 and 7). Furthermore, the high temperature $V_{0.89}Nb_{0.11}O_2$ scattering data shows relatively stronger peaks at odd $L$ in the [0.1 2.9 $L$] cut, and the $A_M \geq 1$ share this scattering behavior. Since the common structural feature linking the different models is the pseudodimer formation induced by elongated apical Nb–O bonds, it follows that the motif is present in high temperature $V_{0.89}Nb_{0.11}O_2$. It is notable that the +$L$ peak in the HT $Nb_{0.11}V_{0.89}O_2$ scattering data is significantly more intense than any of the models reported. This might be addressed by further optimization of the models to favor pseudodimer formation without disrupting the displacement correlation from metal-oxygen size effects. This might be accomplished by including $R_{MM}$ interactions that are shorter than the experimental values (to make the pseudodimers more bond-like), tuning relative $\varepsilon$ values, or by introducing substitutional ordering of aliovalent ions.

In LT $Nb_{0.11}V_{0.89}O_2$, LT $Mo_{0.17}V_{0.83}O_2$, and HT $Mo_{0.17}V_{0.83}O_2$, M1- and M2-like dimerizing displacements are present instead, observable in the intensity around the L/2 planes resulting in extended 2DM2 type ADP correlations, 3D correlated M1 type displacements, and short-range 2DM2 type ADP correlations, respectively. In these structures, dimerizing displacements are strongly coupled to buckling displacements in neighboring chains within the [110] and [-110] planes, whereas dimerizing displacements in HT $Nb_{0.11}V_{0.89}O_2$ are apparently correlated over much shorter ranges and arise from accommodating increased apical Nb–O bond length.

Figure 9 summarizes the assignment of diffuse scattering features (panel A) to distortion types (panel B). The comparison makes clear a key difference between the pseudodimer feature observed in HT $V_{0.89}Nb_{0.11}O_2$ and the metal-metal dimers observed in the other phases. The right portion of Figure 9B shows that the long-short correlations induced by size effects are symmetric around the apical Nb–O bond, which lies on the [110]$_R$ axis. The M1 and related structures are necessarily antisymmetric in their displacements along the same axis, which means that the two displacement networks are of competing symmetries. This is helpful in understanding the regions of reciprocal space of each type of feature. The M1-type broken symmetry is associated with the reciprocal space special point $\mathbf{k} = [½, 0, ½]$ and is seen in the $L/2$ planes (Figure 9A, first 3 panels), while the pseudodimerizing strain field is locally symmetric about the defect, and so the corresponding diffraction feature originates from an integer point (Figure 9B, rightmost panel).

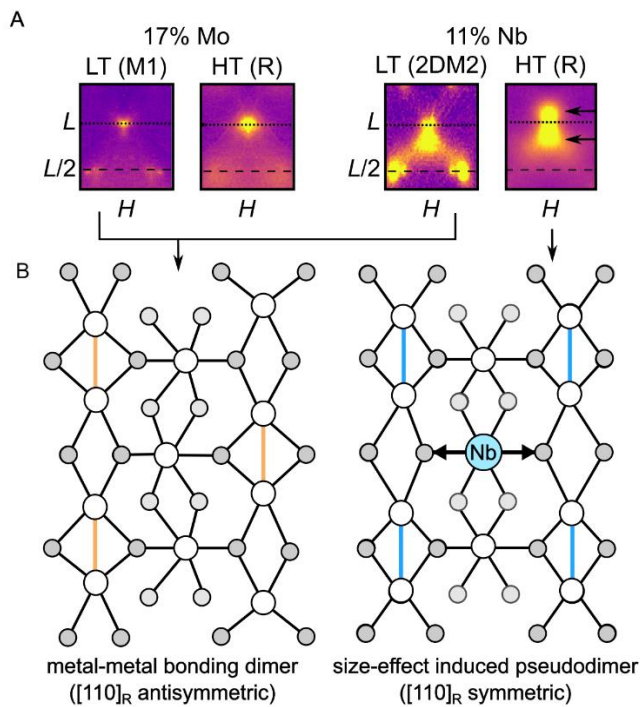

Figure 9: A. Summary of the principle observed diffuse scattering features in $Mo_{0.17}V_{0.83}O_2$ and $Nb_{0.11}V_{0.89}O_2$. B. Comparison of the tiling pattern of the metal-metal dimer phase characteristic of M1 and related distorted rutile phases (left) and the strain-induced pseudodimer (right). The pseudodimers are shown as bonds on the right and the orthogonal dimer network is omitted on the left side for clarity. The arrows between panels indicate the displacement type associated with the observed scattering.

This analysis explains why the double peak feature is not observed in any of the other systems even though the same apical $M$–O bond is expected. Presumably the size effect is overridden by the much stronger metal-metal interactions that generate the longer range dimer formation. That is, the pseudodimer motif only appears in the absence of true metal-metal bonding correlations. This makes the associated size effect diffraction feature a unique example of an observable contraindicator against local dimer formation in this and related examples of doped $VO_2$. In other words, these findings show that an observation of the double size effects peak can be taken as evidence against local dimer formation.

The response of the HT rutile lattice parameters for the Mo and Nb dopant systems can be accounted for by considering their differing contributions to dimer formation in the overall structure. Nb-doping produces an approximately linear increase in $c$ up to about 15% Nb, which is consistent with the pseudodimer strain model shown here. The model does not directly explain the increasingly non-linear composition dependence of $c$ at higher compositions, which implies that more complex interactions become important beyond the dilute regime. Mo-doping introduces essentially the same set of size considerations as Nb-doping does, but it also introduces another electron that favors dimer formation. Thus, the $c$ lattice parameter shows a very shallow parabolic response to Mo-doping because of the competition between two factors: an increase in $c$ due to the increased size of Mo relative to V and enhanced dimer formation due to the extra valence electron of Mo. The equivalent and linear $a$ lattice parameter response in both systems is consistent with the average apical $M$–O bond length increasing with ion size.

It has also been suggested that for some compositions Nb exists as $Nb^{5+}$ ions in Nb-doped $VO_2$ and Mo exists as $Mo^{6+}$ ions in Mo-doped $VO_2$, each with a $d^0$ electron configuration (Villeneuve *et al.*, 1972, Holman *et al.*, 2009, Hiroi, 2015, Lv *et al.*, 2018). In this work the size effects from $Nb^{5+}$ in HT $Nb_{0.11}V_{0.89}O_2$ were modeled in a similar fashion to the models given previously, with Nb having all Nb–O bond lengths as 2.00 Å and with an isotropic octahedral shape. The simulated scattering was not very sensitive to this change (see SI), except for the absence of the +$L$ portion of the double peak feature near [0.0 2.9 3.0], which arises from the less pronounced increases in apical Nb–O bond length relative to the average structure. Thus, if Nb is in the $Nb^{5+}$ oxidation state, a dramatically anisotropic distortion is required to sufficiently alter the apical Nb-O bond necessary to induce pseudodimers and match experiment. Alternatively, even more complex models that introducing correlated $V^{3+}$-$Nb^{5+}$ pairs might produce similar strain fields, but it is difficult to predict what other scattering features would arise from including extra correlations. Thus, treating Nb as $Nb^{4+}$ is the simplest choice for the relatively small 11% substitution level shown here and is also in agreement with recent review of the relevant literature (Pouget, 2021).

Past review of rutile and distorted rutile compounds has emphasized the interplay between the electrostatics of the metal octahedra and the dimerizing interaction between nn metal atoms is fundamental to the understanding of all rutile and distorted rutile compounds (Hiroi, 2015, Baur, 2007), but the discussion has necessarily been focused on crystallographic models based on the average structure This work shows that even something as simple as ion size can induce local structures quite unlike the structure of the end-member compounds being alloyed. Further investigation of these local motifs is required to understand the underlying interactions that produce the varied multitude of distorted rutile structures.

Future studies will aim to account for the total scattering in doped $VO_2$ single crystals by adding ADP correlations to the size effect model described in this study. Inclusion of ADP correlations will account for longer range ordering in the crystals and will allow the model to be extended to compositions beyond the dilute dopant regime. This model will be refined against experimental X-ray scattering for Nb and Mo-doped single crystals to optimize the size effect and ADP correlation parameters for a given composition. These optimized parameters from these refinements will serve as a means of extracting detailed local structure information about these crystals.

## Conclusion

Size effect scattering was observed in synchrotron X-ray diffraction data from $Nb_{0.11}V_{0.89}O_2$ and $Mo_{0.17}V_{0.83}O_2$ single crystals at $T = 100$ and 293 K. Size effect scattering in $L = 1.9$ plane was identical for all 4 phases, but key differences in scattering in the $K = 2.9$ plane were observed. HT $Nb_{0.11}V_{0.89}O_2$ has a unique 'size-effect-like' double peak feature around [0.0 2.9 3.0], and the $+L$ portion of this feature vanishes at low temperature. $Mo_{0.17}V_{0.83}O_2$ shows less pronounced size effect scattering along $L$, with a single peak centered at [0.0 2.9 3.0] at both low and high temperature. A series of size effect models were produced for $Nb_{0.11}V_{0.89}O_2$ in the rutile phase, and it was determined that the double peak scattering motif from HT $Nb_{0.11}V_{0.89}O_2$ is consistent with pseudodimer formation in the rutile chains adjacent to the dopant. The pseudodimers are induced in the host lattice to accommodate the increased Nb−O apical bond length relative to the average structure, and only form when metal-metal interactions are weak. This short-range induced dimerization is unique to HT $Nb_{0.11}V_{0.89}O_2$, whereas metal-metal bonding driven dimerizing displacements in LT $Nb_{0.11}V_{0.89}O_2$, LT $Mo_{0.17}V_{0.83}O_2$, and HT $Mo_{0.17}V_{0.83}O_2$ compete with the pseudodimer formation due to their incompatible symmetries. Future work will extend this local structure model beyond the dilute dopant regime, by accounting for longer range ordering via inclusion of ADP correlations.

**Acknowledgements**

The work was supported by the U.S. Department of Energy, Office of Science, Office of Basic Energy Sciences, EPSCoR and Neutron Scattering Sciences under award DE-SC0018174, U.S. Department of Energy, Office of Science, Materials Sciences, and Engineering Division and Scientific User Facilities Division. Use of the Advanced Photon Source at Argonne National Laboratory was supported by the U. S. Department of Energy, Office of Science, Office of Basic Energy Sciences, under Contract No. DE-AC02-06CH11357.

# Supporting information for Competition between metal bonding and strain in tetragonal V$_{1-x}$$M_x$O$_2$ ($M$ = Nb, Mo)

Jacob F. Phillips[1], Tyra C. Douglas[1], Matthew A. Davenport[1], Top B. Rawot Chhetri[1], Logan M. Whitt[1], Stephan Rosenkranz[2], Raymond Osborn[2], Matthew J. Krogstad[3], and Jared M. Allred[1]

1. Department of Chemistry and Biochemistry, University of Alabama, Tuscaloosa, AL 35487
2. Materials Science Division, Argonne National Laboratory, Lemont, IL
3. Advanced Photon Source, Argonne National Laboratory, Lemont, IL, 60439

## MONTE CARLO SIMULATION DETAILS

In each MC simulation, 1000 $N$ MC cycles were performed, where $N$ is the number of atoms in the simulated crystal. The random atomic displacements were allowed to take values within a Gaussian distribution with a width of 0.005 along each average crystal lattice fractional coordinate. Energetically favorable moves are always accepted, and energetically unfavorable moves are accepted based on the change in total energy, which is calculated using 12–6 Lennard-Jones (LJ) potentials, with a move acceptance probability, $P = \frac{e^{-\Delta E'}}{1+ e^{-\Delta E'}}$, where $\Delta E' = \frac{\Delta E}{kT}$.

## DISCUS X-RAY SCATTERING SIMULATION DETAILS

For each simulation, the scattering was simulated in reciprocal space planes with step sizes of 0.05 reciprocal lattice units along each dimension. The X-ray wavelength used for the simulations was 0.1425 nm. The scattering was calculated for 150 randomly selected 20 × 20 × 20 unit cell portions of the model crystal and summed to produce the scattering map. The scattering from the average structure was calculated from 50% of the model crystal and was subtracted from the total scattering to yield the simulated X-ray diffuse scattering. The resulting scattering map was then symmetrized according to its Laue group *4/mmm*.

**Table S1: Complete set of pairs used in simulations Nb models.**

| Pair Type | Name |
|---|---|
| $(V–O)_1$; $(Nb–O)_1$ | Apical $M$–O |
| $(V–O)_2$; $(Nb–O)_2$ | Equatorial $M$–O |
| $(V–V)$ | |
| $(V–Nb)$ | Intrachain $M$–$M'$ pair |
| $(Nb–Nb)$ | |

**Table S2: List of reduced parameters for referencing model types, with their expression in base LJ terms**

| Symbol | Expression | Parameter |
|---|---|---|
| $\sigma'(i–j)$ | $\sqrt[6]{2}\sigma(i–j)$ | LJ target ($i$–$j$) distance |
| $\bar{\sigma}'(M–O)$ | $[\sigma'(M-O)_1 + 2\,\sigma'(M-O)_2]/3$ | Average $\sigma'(M$–O$)$ |
| $\alpha_M$ | $\sigma'(M–O)_1/\sigma'(M–O)_2$ | Target ($M$–O) anisotropy |
| $R_{MO}$ | $\bar{\sigma}'(Nb–O)/\bar{\sigma}'(V–O)$ | Ratio of Nb–O and V–O target bond length averages |
| $A_M$ | $\alpha_{Nb}/\alpha_V$ | Anisotropy parameter ratio |
| $R_{MM}$ | $\sigma'(Nb–Nb)\ \sigma'(V–V)$ | Difference between Nb–Nb and V–V target distances |

**Table S2: Constrained LJ parameters from chosen model parameters: $R_{M–O}$ and $A_M$.**

| Model | | Constrained | | | | | |
|---|---|---|---|---|---|---|---|
| $R_{M–O}$ | $A_M$ | $\alpha_{Nb}$ | $\alpha_V$ | $\sigma'(V–O)_1$ | $\sigma'(V–O)_2$ | $\sigma'(Nb–O)_1$ | $\sigma'(Nb–O)_2$ |
| 1.061 | 1.000 | 0.996 | 0.996 | 1.923 Å | 1.931 Å | 2.040 Å | 2.048 Å |
| 1.057 | 0.955 | 0.957 | 1.001 | 1.931 Å | 1.929 Å | 1.980 Å | 2.070 Å |
| 1.057 | 1.056 | 1.045 | 0.990 | 1.916 Å | 1.936 Å | 2.100 Å | 2.010 Å |

**Table S3: Constrained LJ metal-metal parameters from chosen $R_{M–M}$.**

| Model | Constrained | | |
|---|---|---|---|
| $R_{M–M}$ | $\sigma'(V–V)$ | $\sigma'(V–Nb)$ | $\sigma'(Nb–Nb)$ |
| 1.045 | 2.879 Å | 2.944 Å | 3.009 Å |
| 0.955 | 2.908 Å | 2.843 Å | 2.778 Å |

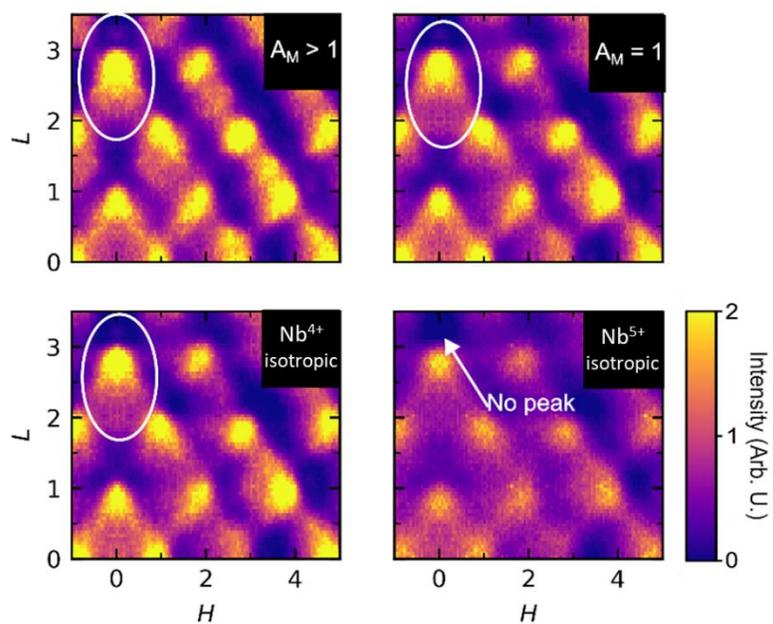

Figure S1: Comparison of simulated X-ray diffuse scattering from a model with Nb in the 4+ oxidation state ($A_M > 1$, $A_M = 1$, $Nb^{4+}$ isotropic) to a model with Nb in the 5+ oxidation state ($Nb^{5+}$ isotropic).

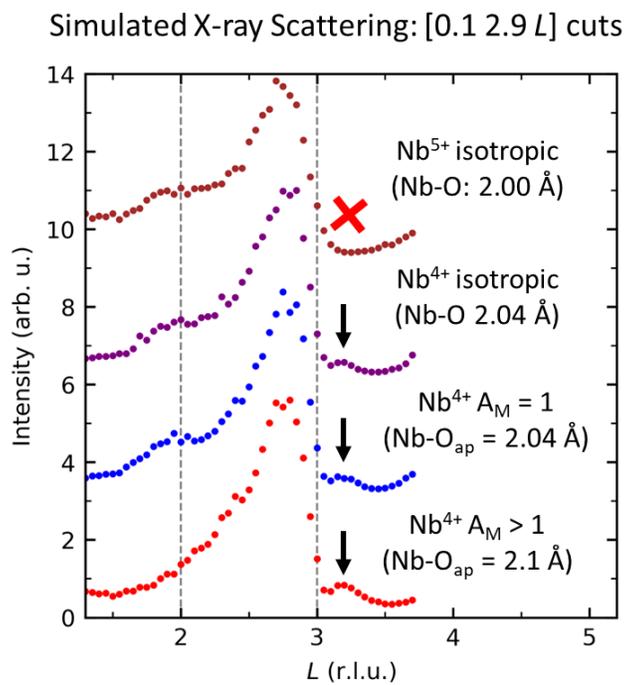

Figure S2: [0.1 2.9 $L$] cuts of simulated scattering from selected models. Note the presence of a peak on the $+L$ side of [0.1 2.9 3.0] for models with increased apical Nb–O bond length and the absence of this peak for the $Nb^{5+}$ model.